\documentclass[12pt,amsfonts]{article}
\begin{document}
\def\p {{\partial}}
\def\n {{\nu}}
\def\m {{\mu}}
\def\a {{\alpha}}
\def\bt {{\beta}}
\def\f {{\phi}}
\def\th {{\theta}}
\def\g {{\gamma}}
\def\eps {{\epsilon}}
\def\e {{\psi}}
\def\la {{\lambda}}
\def\na {{\nabla}}
\def\k {\chi}
\def\bn {\begin{eqnarray}}
\def\en {\end{eqnarray}}
\title{Formulation of Hamiltonian equations for fractional variational problems} \maketitle
\begin{center}
\author{\textbf{Sami I. Muslih}
\vspace{0.2cm}\footnote{E-mail:
smuslih@ictp.trieste.it}\\Department
of Physics, Al-Azhar University, Gaza, Palestine\\and\\
International Center for Theoretical Physics(ICTP),\\Trieste,
Italy\\
\vspace{1cm} \textbf{Dumitru Baleanu} \vspace{0.2cm}\footnote{On
leave of absence from Institute of Space Sciences, P.O.BOX, MG-23,
R 76900, Magurele-Bucharest, Romania,E-mails:
dumitru@cankaya.edu.tr,
baleanu@venus.nipne.ro}\\
Department of Mathematics and Computer Sciences, Faculty of Arts
and Sciences,
\c{C}ankaya University- 06530, Ankara, Turkey }\\
\end{center}
\hskip 5 cm
\begin{abstract}
 An extension of Riewe's fractional Hamiltonian formulation is presented for fractional constrained systems. The conditions of
consistency of the set of constraints with equations of motion are
investigated. Three examples of fractional constrained systems are
analyzed
in details.
\end{abstract}

\newpage

\section{Introduction}

Fractional derivatives [1-6], have played a significant role in
engineering, science, and pure and applied mathematics during the
last decade [7-10]. Fractional calculus found many applications in
recent studies of scaling phenomena [11-13]. One can find other
important and interesting application of fractional calculus in
classical mechanics [14-20]. Riewe has used the fractional
calculus to develop a formalism which can be used for both
conservative and non conservative systems [14, 15]. Although many
laws of nature can be obtained using certain functionals and the
theory of calculus of variations, not all laws can be obtained by
this way. For example, almost all systems contain internal
damping, yet traditional energy based approach cannot be used to
obtain equations describing the behavior of a non conservative
system [14,15].By this approach, one can obtain the Lagrangian and
the Hamiltonian equations of motion for the nonconservative
systems .
\\
Klimek have introduced the symmetric fractional derivative and the
Euler- Lagrange equations for models depending on sequential
derivatives were obtained by using the minimal action principle
[9]. In addition, the correspond- ing Hamiltonian was introduced
and the Hamilton's equations were analyzed [9]. The models
described by fractional order derivatives of Riemann-Liouville
type in sequential form were analyzed in both Lagrangian and
Hamiltonian formalisms in [10].
\\
Taking into account the constrained systems play an important role
in gauge theories and the distinguishing feature of singular
Lagrangian theories in the Hamiltonian formulation is the presence
of constraints, we believe that the application of fractional
calculus to constrained systems presents interest both from
theoretical and practical point of views. Recently, an extension
of the simplest fractional problem and the fractional variational
problem of La- grange was obtained [18,19]. A natural
generalization of Agrawal's approach [18,19], is to apply the
fractional calculus to constrained systems [21-22] and to obtain
the Hamiltonian formulation of constrained systems.

The main aim of this paper is to obtain the Hamiltonian equations
of motion for fractional variational problems with constraints.

The plan of this paper is as follows: \\
In Sect. 2
Riemnann-Liouville fractional derivatives are briefly reviewed.
Sect. 3 is dealing with fractional Hamiltonian equations initiated
by Riewe. Sect. 4 presents the Agrawal's approach of the
fractional Euler-Lagrange equations. In Sect. 5 the extension of
Riewe's formulation for the constrained systems is obtained. Three
examples of fractional constrained systems are investigated in
Sect. 6 . Finally, Sect. 7 is dedicated to our conclusions.

\section{Riemann-Liouville fractional derivative}

Several definitions of a fractional derivative has been proposed.
These definitions include Riemann-Liouville, Grunwald-Letnikov,
Weyl, Caputo, Marchaud and Riesz fractional derivatives [1-6].
Here, we formulate the problem in terms of the left and right
Riemann-Liouville fractional derivatives, which are defined as [5]

${\it The~ left~ Riemann-Liouville~ fractional~ derivative}$

\begin{equation}
{{}_a\textbf{D}_t^{\alpha}f(t)} =
\frac{1}{\Gamma{(n-\alpha)}}\left(\frac{d}{dt}\right)^{n}\int\limits_a^t(t-\tau)^{n-\alpha-1}f(\tau)d\tau,
\end{equation}

and

${\it The~ right ~Riemann-Liouville ~fractional~ derivative}$

\begin{equation}
{{}_t\textbf{D}_b^{\alpha}f(t)} =
\frac{1}{\Gamma{(n-\alpha)}}\left(-\frac{d}{dt}\right)^{n}\int\limits_t^b(\tau-t)^{n-\alpha-1}f(\tau)d\tau,
\end{equation}
where the order $\alpha$ fulfills $n-1\leq\alpha <n$ and $\Gamma$
represents the Euler's gamma function. If $\alpha$ is an integer,
these derivatives are defined in the usual sense, i.e.,
\begin{equation}
{{}_a\textbf{D}_t^{\alpha}f(t)}
=\left(\frac{d}{dt}\right)^{\a},~~{{}_t\textbf{D}_b^{\alpha}f(t)}
= \left(-\frac{d}{dt}\right)^{\a}, ~\a=1,2,... .
\end{equation}

\section{Riewe's formulation}

 Now we shall briefly review Riewe's formulation of
fractional generalization of Lagrangian and Hamiltonian equations
of motion.

Let us consider the action function of the form [14,15]
\begin{equation}
S=\int_{a}^{b} L(\{q_{n}^{r}, Q_{n'}^{r}\}, t)dt,
\end{equation}
where the generalized coordinates are defined as follows:
\begin{equation}
q_{n}^{r}:= {({}_a\textbf{D}_t^{\alpha})^{n}
x_{r}(t)},~Q_{n'}^{r}:= {({}_t\textbf{D}_b^{\alpha})^{n'}
x_{r}(t)},
\end{equation}
and $r=1,2,..., R $ denotes the number of fundamental coordinates,
$ n=0,..., N,$ the sequential order of the derivatives defining
the generalized coordinates $q$, and $n'=1,..., N'$ the sequential
order of the derivatives in definition of the coordinates $Q$. A
necessary condition for $S$ to admit an extremum for given
functions $x_{r}(t)$ is that $x_{r}(t)$ satisfy the Euler-Lagrange
equations [14,15]
\begin{equation}
\frac{\p L}{\p q_{0}^{r}} +
\sum_{n=1}^{N}{({}_t\textbf{D}_b^{\alpha})^{n}}\frac{\p L}{\p
q_{n}^{r}} +
\sum_{n'=1}^{N'}{({}_a\textbf{D}_t^{\alpha})^{n'}}\frac{\p L}{\p
Q_{n'}^{r}}  =0.
\end{equation}

Following references [14,15], the generalized momenta take the
form \bn && p_{n}^{r} =
\sum_{k=n+1}^{N}{({}_t\textbf{D}_b^{\alpha})^{k-n-1}}\frac{\p
L}{\p q_{k}^{r}},\nonumber\\
&& \pi_{n'}^{r} =
\sum_{k=n'+1}^{N'}{({}_a\textbf{D}_t^{\alpha})^{k-n'-1}}\frac{\p
L}{\p Q_{k}^{r}}. \en

The canonical Hamiltonian has the following form:
\begin{equation}
H = \sum_{r=1}^{R}\sum_{n=0}^{N-1} p_{n}^{r}q_{n+1}^{r} +
\sum_{r=1}^{R}\sum_{n'=0}^{N'-1} \pi_{n'}^{r}Q_{n'+1}^{r} - L.
\end{equation}

The Hamilton's  equations of motion read as
\begin{equation}
\frac{\p H}{\p q_{N}^{r}}=0,~\frac{\p H}{\p Q_{N'}^{r}}=0.
\end{equation}
For $n=1,..., N,~n'=1,..., N'$ we have the following equations of
motion

\bn &&\frac{\p H}{\p q_{n}^{r}}= {{}_t\textbf{D}_b^{\alpha}
p_{n}^{r}},~~\frac{\p H}{\p Q_{n'}^{r}}=
{{}_a\textbf{D}_t^{\alpha} \pi_{n'}^{r}},\\
&&\frac{\p H}{\p q_{0}^{r}} = - \frac{\p L}{\p q_{0}^{r}} =
{{}_t\textbf{D}_b^{\alpha} p_{0}^{r}} + {{}_a\textbf{D}_t^{\alpha}
\pi_{0}^{r}}.\en

The other equations are given by \bn &&\frac{\p H}{\p
p_{n}^{r}}=q_{n+1}^{r}= {{}_a\textbf{D}_t^{\alpha} q_{n}^{r}},
~~\frac{\p H}{\p \pi_{n'}^{r}}=Q_{n+1}^{r}=
{{}_t\textbf{D}_b^{\alpha}
Q_{n'}^{r}}, \\
&&\frac{\p H}{\p t} = - \frac{\p L}{\p t}, \en where, $n=0,...,
N,~n'=1,..., N'$.

\section{Agrawal's approach}

Agrawal have obtained  the Euler-Lagrange equations for fractional
variational problems [18]. Now we would like to give review for
his approach.

Consider the action function
\begin{equation}
S[q_{0}^{1},....q_{0}^{R}] =\int_{a}^{b}L(\{q_{n}^{r}, Q_{n'}^{r}\}, t)dt,
\end{equation}

subject to the independent constraints
\begin{equation}
\Phi_{m}(t, q_{0}^{1},..., q_{0}^{R}, q_{n}^{r},Q_{n'}^{r})=0, ~m < R,
\end{equation}
where the generalized coordinates are defined as follows:
\begin{equation}
q_{n}^{r}:= {({}_a\textbf{D}_t^{\alpha})^{n}
x_{r}(t)},~Q_{n'}^{r}:= {({}_t\textbf{D}_b^{\bt})^{n'} x_{r}(t)},
\end{equation}

Then the necessary condition for the curves $q_{0}^{1},....,
q_{0}^{R}$ with the boundary conditions
$q_{0}^{r}(a)=q_{0}^{ra},~q_{0}^{r}(b)=q_{0}^{rb},r=1, 2,,...,R$,
to be an extremal of the functional given by equation (14) is that
the functions $q_{0}^{r}$ satisfy the following Euler-Lagrange
equations [18]:
\begin{equation}
\frac{\p \bar{L}}{\p q_{0}^{r}} +
\sum_{n=1}^{N}{({}_t\textbf{D}_b^{\alpha})^{n}}\frac{\p \bar
{L}}{\p q_{n}^{r}} +
\sum_{n'=1}^{N'}{({}_a\textbf{D}_t^{\a})^{n'}}\frac{\p \bar{L}}{\p
Q_{n'}^{r}}  =0,
\end{equation}

where $\bar{L}$ is defined as [18]
\begin{equation}
\bar{L}(\{q_{n}^{r}, Q_{n'}^{r}\}, t, \lambda_{m}(t)) =
L(\{q_{n}^{r}, Q_{n'}^{r}\}, t) + \lambda_{m}(t)\Phi_{m}(t,
q_{0}^{1},..., q_{0}^{R}, q_{n}^{r},Q_{n'}^{r}),
\end{equation}
here, the multiple $\lambda_{m}(t)\in R^{m}$ are continuous on
$[a, b]$.

\section{Fractional Hamiltonian formulation}

In fractional calculus it is not a unique way to define the
Hailtonian for a given Lagrangian, mainly because they are several
definitions of fractional derivatives.

 To obtain the Hamilton's
equations for the the fractional variational problems, we  need to
re-define the lift and the right canonical momenta as follows: \bn
&& p_{n}^{r} =
\sum_{k=n+1}^{N}{({}_t\textbf{D}_b^{\alpha})^{k-n-1}}\frac{\p
\bar{L}}{\p q_{k}^{r}},\nonumber\\
&& \pi_{n'}^{r} =
\sum_{k=n'+1}^{N'}{({}_a\textbf{D}_t^{\a})^{k-n'-1}}\frac{\p
\bar{L}}{\p Q_{k}^{r}}. \en

The canonical Hamiltonian has the following form:
\begin{equation}
{\bar{H}} = \sum_{r=1}^{R}\sum_{n=0}^{N-1} p_{n}^{r}q_{n+1}^{r} +
\sum_{r=1}^{R}\sum_{n'=0}^{N'-1} \pi_{n'}^{r}Q_{n'+1}^{r} -
{\bar{L}}.
\end{equation}
Then, the modified canonical equations of motion are obtained as
\bn&&\{q_{n}^{r}, \bar{H}\}= {{}_t\textbf{D}_b^{\alpha}
p_{n}^{r}},~\{Q_{n'}^{r}, \bar{H}\} ={{}_a\textbf{D}_t^{\alpha} \pi_{n'}^{r}},\\
&&\{q_{0}^{r}, \bar{H}\} = {{}_t\textbf{D}_b^{\alpha} p_{0}^{r}} +
{{}_a\textbf{D}_t^{\alpha} \pi_{0}^{r}} .\en Here, $n=1,...,
N,~n'=1,..., N'$.

The other set of equations of motion are obtained as \bn &&
\{p_{n}^{r}, \bar{H}\} =q_{n+1}^{r}= {{}_a\textbf{D}_t^{\alpha}
q_{n}^{r}},~\{\pi_{n'}^{r}, \bar{H}\}\}= Q_{n+1}^{r}=
{{}_t\textbf{D}_b^{\alpha}
Q_{n'}^{r}}, \\
&&\frac{\p \bar{H}}{\p t} = - \frac{\p \bar{L}}{\p t}. \en Here,
$n=0,..., N,~n'=1,..., N'$ and the commutator $\{, \}$ is the
Poisson's bracket and it is defined as
\begin{equation}
\{A, B\}_{q_{n}^{r}, p_{n}^{r}, Q_{n'}^{r}, \pi_{n'}^{r}} =
\frac{\p A}{\p q_{n}^{r}}\frac{\p B}{\p p_{n}^{r}}- \frac{\p B}{\p
q_{n}^{r}}\frac{\p A}{\p p_{n}^{r}} + \frac{\p A}{\p
Q_{n'}^{r}}\frac{\p B}{\p \pi_{n'}^{r}}- \frac{\p B}{\p
Q_{n'}^{r}}\frac{\p A}{\p \pi_{n'}^{r}},
\end{equation}
where, $n=0,..., N,~n'=1,..., N'$.

\section{Examples}
In this section, we obtain the Hamiltonian equations for an
constrained fractional variational problems. All calculations were
done for $0<\alpha < 1$.

\subsection{Example 1}

As the first example, consider the following constrained
fractional variational problem from the optimal control theory
[18]: minimize
\begin{equation}
S[x_{1}, x_{2}] =\frac{1}{2}\int_{0}^{1} {[ x_{1}^{2} +
x_{2}^{2}]}dt,
\end{equation}
such that \bn&& {{}_0\textbf{D}_t^{\alpha} x_{1}}= - x_{1} +
x_{2},\\
&&x_{1}(0)= 1.
\en

The modified Lagrangian $\bar{L}$ is given by
\begin{equation}
\bar{L} = \frac{1}{2}{[ x_{1}^{2} + x_{2}^{2}]} + l\phi_{1} +
{\la} \phi_{2},
\end{equation}
where $l, {\la} $ are Lagrange multipliers and $\phi_{1},
\phi_{2}$ are the constraints and are defined as \bn && \phi_{1} =
{{}_0\textbf{D}_t^{\alpha} x_{1}} + x_{1} -
x_{2}= 0,\\
&&\phi_{2} = x_{1}(0) -1 =0.
\en

The generalized canonical momenta read as
\begin{equation}
p_{0}^{1} = l, ~~~p_{0}^{2} =0.
\end{equation}

The  Hamiltonian $\bar {H}$ is given by
\begin{equation}
\bar{H} = p_{0}^{1} q_{1}^{1} - \frac{1}{2}{[ x_{1}^{2} +
x_{2}^{2}]} - l\phi_{1} - {\la} \phi_{2},
\end{equation}

The Hamiltonian equations are obtained as
\begin{equation}
\frac{\p \bar{H}}{\p q_{1}^{1}} = p_{0}^{1} - l=0,
\end{equation}
\begin{equation}
\frac{\p \bar{H}}{\p q_{1}^{2}} =0,
\end{equation}
which is identically satisfied. The other equations of motion are
given by \bn &&\frac{\p \bar{H}}{\p x_{1}} = -x_{1} - l =
{{}_t\textbf{D}_1^{\alpha}(p_{0}^{1}}),\\
&&\frac{\p \bar{H}}{\p x_{2}}=-x_{2} + l = 0. \en

Making use of (33), equations (36) and (37) can be put in the
form, \bn
&&x_{1} + l + {{}_t\textbf{D}_1^{\alpha}(l)} = 0,\\
&& x_{2} - l = 0.\en

Equations (38) and (39) are in exact agreement with those obtained
by using the Euler-Lagrange formulation for fractional variational
problems [18].

\subsection{Example 2}

Let us consider the following Lagrangian

\begin{equation}\label{uu}
L=\frac{1}{2}({\dot x_1}+{\dot x_2})^2.
\end{equation}
The Lagrangian (\ref{uu})is singular because its usual Hessian
matrix has rank 1. One possible fractional generalization of
(\ref{uu}) is given as follows

\begin{equation}\label{uuu}
{\bar L}= \frac{1}{2}({}_0\textbf{D}_t^{\alpha} x_1+
{}_0\textbf{D}_t^{\alpha} x_2)^2.
\end{equation}
For Lagrangian (\ref{uuu}) we introduce the notion of fractional
Hessian as
\begin{equation}\label{hes}
\frac{\partial^2 {\bar L}}{\partial {}_0\textbf{D}_t^{\alpha} x_i
{}_0\textbf{D}_t^{\alpha} x_j }, i,j=1,2.
\end{equation}
Evaluating the rank of (\ref{hes}) we obtain the result being 1,
so the system is a constrained system.

 On the other hand from (\ref{uuu}) we obtain
immediately the form of the canonical momenta as

\begin{equation}\label{uuuu}
p_0^1={{}_0\textbf{D}_t^{\alpha} x_1} +{{}_0\textbf{D}_t^{\alpha}
x_2},p_0^2={{}_0\textbf{D}_t^{\alpha} x_1}
+{{}_0\textbf{D}_t^{\alpha} x_2},
\end{equation}

so that (\ref{uuuu}) implies that

\begin{equation}\label{const}
p_0^2-p_0^1=0,
\end{equation}

which represents a primary constraint.

Using (\ref{uuu}) and (\ref{uuuu}) the form of the canonical
fractional Hamiltonian is given by

\begin{equation}\label{hah}
{\bar H}=\frac{1}{2}(p_0^1)^2 + {}_0\textbf{D}_t^{\alpha}
x_2(p_0^2-p_0^1).
\end{equation}

If we denote in (\ref{hah}) ${}_0\textbf{D}_t^{\alpha} x_2$ by
$\lambda$ we observe that canonical fractional Hamiltonian becomes

\begin{equation}\label{h}
{\bar H}=\frac{1}{2}(p_0^1)^2 +\lambda (p_0^2-p_0^1).
\end{equation}
From (\ref{h}) we obtain the fractional Hamiltonian motions as
follows

\begin{equation}
{}_0\textbf{D}_t^{\alpha} x_1=p_0^1-\lambda,
{}_0\textbf{D}_t^{\alpha} x_2=\lambda. \end{equation}
\begin{equation}\label{ecu}
 {}_0\textbf{D}_t^{\alpha}
p_0^1=0,{}_0\textbf{D}_t^{\alpha} p_0^2=0.
\end{equation}

Using (\ref{const}) and (\ref{ecu}) we conclude that the primary
surface of constraints is preserved after considering its
evolution from the fractional calculus point of view.

\subsection{Example 3}

Finally, let us analyze the following Lagrangian

\begin{equation}
L=\frac{1}{2}(\dot x_1)^2-{\dot x_2}x_3.
\end{equation}
One possible generalization of (49) is given as
\begin{equation}\label{l2}
{\bar L}=\frac{1}{2}({}_0\textbf{D}_t^{\alpha} x_1)^2
-({{}_0\textbf{D}_t^{\alpha} x_2})x_3.
\end{equation}

The fractional Hessian corresponding to (50) is given by

\begin{equation}\label{hes}
\frac{\partial^2 {\bar L}}{\partial {}_0\textbf{D}_t^{\alpha} x_i
{}_0\textbf{D}_t^{\alpha} x_j }, i,j=1,3.
\end{equation}
has rank one, therefore the fractional Lagrangian (\ref{l2}) is
degenerate admitting two primary constraints.

 Using (\ref{l2}) we
obtain the momenta as follows

\begin{equation}\label{l2m}
p_0^1={}_0\textbf{D}_t^{\alpha} x_1, p_0^2=x_3,p_0^3=0.
\end{equation}

From (52) we identify the primary constraints as

\begin{equation}
p_0^2-x_3=0,p_0^3=0.
\end{equation}

The canonical Hamiltonian in this case is given by

\begin{equation}\label{l2h}
{\bar H}=\frac{(p_0^1)^2}{2} +{{}_0\textbf{D}_t^{\alpha}
x_2}(p_0^2-x_3).
\end{equation}

Denoting ${{}_0\textbf{D}_t^{\alpha} x_2}$ by $\lambda$ and by
using (\ref{l2h}) the canonical equations become

\begin{eqnarray}\label{l2eq}
{}_0\textbf{D}_t^{\alpha} x_1&=&p_0^1,{}_0\textbf{D}_t^{\alpha}
x_3=0,{}_0\textbf{D}_t^{\alpha} x_2=\lambda,\cr
{}_0\textbf{D}_t^{\alpha} p_0^1&=&0,{}_0\textbf{D}_t^{\alpha}
p_0^3=-\lambda, {}_0\textbf{D}_t^{\alpha} p_0^2=0,p_0^2-q_3=0.
\end{eqnarray}
We observed from (55) that the consistency conditions of the
primary constraint $p_0^2-x_3=0$ is compatible with the equations
of motion.

The evolution of the second primary constraint makes $\lambda$
zero, therefore

\begin{equation}
{}_0\textbf{D}_t^{\alpha} x_2=0.
\end{equation}

From (56) we conclude that a new secondary constraint appears,
therefore the initial phase-space was modified.

\section{Conclusion}

The fractional quantization problem is an open and important
issue. The main difficulty is related to the non-locality of
fractional Lagrangian. Another specific problem in this field is
related to the existence of various definitions of fractional
derivatives which lead to different forms of fractional
Lagrangians and fractional Hamiltonians.

In this study, the Hamiltonian equations have been obtained for
unconstrained and constrained variational problem, in the same
manner as those obtained by using the formulation of
Euler-Lagrange equations for variational problems obtained in
[18].  The Hamiltonian describes a non-conservative system,
therefore the the Poisson brackets introduced in this study leads
us to a Hamiltonian conserved in time in the limit
$\alpha\rightarrow 1$ and for Hamiltonian which does not
explicitly depends on time. In this paper two different systems
with constraints were investigated, the first kind corresponds to
a fractional Lagrangian containing Lagrange multipliers. For this
case the fractional Hamiltonian and its corresponding equations
were obtained.

The second kind of constrained system involves primary and
secondary constraints. Having in mind to obtain similar results
with the usual analysis of constrained systems in the limit
process we investigated the form of the equations of motion and
the evolution of the surface of primary constraints for the last
two examples. It was observed that the form of the primary
constraints and the form of the Hamiltonian equations are in
agreement with each other for the second example but for the third
one the consistency conditions impose new constraints.

In the general case the primary constraints may produce secondary
constraints and the process may continue until no new constraints
will appear.

\section*{Acknowledgments}

S. M. would like to thank the Abdus Salam International Center for
Theoretical Physics, Trieste, Italy, for support and hospitality
during the preparation of this work. This work was done within the
framework of the Associateship Scheme of the Abdus Salam ICTP.\\
D. B.  would like to thank  O. Agrawal and J. A. Tenreiro Machado
for interesting discussions and to K. B. Oldham for
encouragements. This work is partially supported by the Scientific
and Technical Research Council of Turkey.

\end{document}